\documentclass[11pt, a4paper]{article}
\usepackage{moriond,epsfig}

\usepackage{wrapfig}

\def\be{\begin{equation}}
\def\ee{\end{equation}}
\def\bea{\begin{eqnarray}}
\def\eea{\end{eqnarray}}

\newcommand{\msolar}{{\rm M_{\odot}}}

\newcommand{\temp}{{\rm T}}
\newcommand{\pre}{{\rm P}}
\newcommand{\pram}{\pre_{\mbox{ram}}}
\newcommand{\pth}{{\rm P}_{\mbox{th}}}
\newcommand{\pmag}{{\rm P}_{\mbox{mag}}}
\newcommand{\vrms}{{\rm V}_{{\rm rms}}}

\begin{document}
\vspace*{4cm}
\title{Disk-Halo-Disk Circulation and the Evolution of the ISM - 3D HD and MHD Adaptive Mesh Refinement Simulations}

\author{M.A. de Avillez (1,2), D. Breitschwerdt (2)}
\address{(1) Department of Mathematics, University of \'Evora, R. Rom\~ao Ramalho 59, 7000 \'Evora, Portugal}
\address{(2) Institut f\"ur Astronomie, Universit\"at Wien, T\"urkenschanzstr. 17, A-1180 Wien, Austria}

\maketitle\abstracts{State of the art models of the ISM use adaptive mesh refinement to capture small scale structures, by refining on the fly those regions of the grid where density and pressure gradients occur, keeping at the same time the existing resolution in the other regions. With this technique it became possible to study the ISM in star-forming galaxies in a global way by following matter circulation between stars and the interstellar gas, and, in particular the energy input by random and clustered supernova explosions, which determine the dynamical and chemical evolution of the ISM, and hence of the galaxy as a whole. In this paper we review the conditions for a self-consistent modelling of the ISM and present the results from the latest developments in the 3D HD/MHD global models of the ISM. Special emphasis is put on the effects of the magnetic field with respect to the volume and mass fractions of the different ISM ``phases'', the relative importance of ram, thermal and magnetic pressures, and whether the field can prevent matter transport from the disk into the halo. The simulations were performed on a grid with a square area of 1 kpc$^{2}$, centered on the solar circle, extending $\pm 10$ kpc perpendicular to the galactic disk with a resolution as high as 1.25 pc. The run time scale was 400 Myr, sufficiently long to avoid memory effects of the initial setup, and to allow for a global dynamical equilibrium to be reached in case of a constant energy input rate.} %
\noindent
{\small¥{\it Keywords}: magneto-hydrodynamics -- galaxies: ISM --
galaxies: kinematics
  and dynamics -- Galaxy: disk -- Galaxy: evolution -- ISM: bubbles --
  ISM: general -- ISM: kinematics and dynamics -- ISM: structure}

\section{Introduction}

So far our understanding of the evolution of the ISM has been scanty,
because of the inherent nonlinearity of all the processes
involved. Analytic ISM models, which tried to explain the distribution
of the ISM gas into various thermally stable ``phases'' such as the
pioneering work by, e.g., Field (1965), Field et al. (1969), Cox \&
Smith (1974), and McKee \& Ostriker (1977), can at most be regarded as
exploratory. Noticeable progress was only possible by the development of sophisticated numerical codes, adequate computing power, and precision input data by observations. Only very recently, by the rapid evolution of telescope and detector technology, as well as the availability of large numbers of parallel processors, we are in the fortunate situation to follow in detail the evolution of the ISM on
the global scale taking into account the disk-halo-disk circulation in three dimensions (the first models using 2D grids were developed by Rosen \& Bregman 1995).

In this paper we discusss the most important numerical prerequisites for a realistic and self-consistent modelling of the ISM (Section 2), followed by a brief summary of the latest 3D disk-halo-disk circulation simulations (Section 3). Section 4 discusses model testing and finally, in Section 5 a few final remarks are presented.

\section{Modelling of the ISM}

The key to a realistic description of the ISM is the use of
the apropriate dimensionality, grid coverage, the highest possible
spatial resolution, and a realistic input of the basic physical
processes with appropriate boundary and initial conditions. All this
coupled to the appropriate tools for the included physics, which are
sophisticated HD and MHD codes capable of tracking non-linear and small scale structures, solving the Riemann problem between neighbouring cells without introducing large artificial viscosity and guarantee the conservation of $\nabla\cdot{\rm B}=0$ in case magnetic fields are included. These simulations are considered to provide a reliable description of the ISM providing the memory of the initial conditions and evolution has been lost and the models used in these runs have been tested against observations in the local ISM and in particular within the Local Bubble (Cox 2004; Breitschwerdt \& Cox 2004, see also Breitschwerdt et al. in this volume).
 
{\bf Dimensionality} is of crucial importance as it determines
the dynamics (and the turbulence) of the flow with/without magnetic fields. For example, the idea that a disk parallel magnetic field could suppress break-out and outflow into the halo was mainly based on 2D simulations carried out in the last 15 years, owing to computing power limitations. It is obvious, that in 2D-MHD, the flow perpendicular to the magnetic field lines (and hence to the galactic plane) is subject to opposing magnetic tension forces. In 3D however, field lines can be pushed aside and holes and channels can be punched into the gas and field, allowing pressurized flow to
circumvent \textit{ever increasing tension and pressure forces} in
$z$-direction. 

This behaviour requires that the {\bf grid coverage} in the $z-$direction must
be large enough to accomodate the outflows, whose upper height above the
Galactic plane can be estimated by calculating the flow time, $\tau_f$, that the gas needs to travel to the critical point of the flow in a steady-state (see Kahn 1981). This is the characteristic distance from which information in a thermally driven flow can be communicated back to the sources.  Then
$\tau_f \sim r_c/c_s$, where $r_c$ and $c_s$ are the location of the
critical point and the speed of sound, respectively. For spherical
geometry, the critical point can be simply obtained from the steady
state fluid equations, $r_c \sim G M_{gal}/(2 c_s^2)$, which yields with
a Milky Way mass of $M_{gal} \approx 4 \times 10^{11} \, {\rm
M_{\odot}}$ a distance $r_c \approx 31.3$ kpc for an isothermal gas at
$T= 2 \times 10^6$ K (corresponding to a sound speed of $1.67 \times
10^7 {\rm cm} \, {\rm s}^{-1}$), and thus $\tau_f \sim 180$ Myr as an
upper limit for the flow time. For comparison, the radiative cooling
time of the gas at a typical density of $n=2 \times 10^{-3} \, {\rm
cm}^{-3}$ is roughly $\tau_c \sim 3 k_B T/(n \Lambda) \approx 155$ Myr
for a standard collisional ionization equilibrium cooling function
$\Lambda = 8.5 \times 10^{-23} \, {\rm   erg}\, {\rm cm}^3 \, {\rm
s}^{-1}$ of gas with cosmic abundances. This value is apparently of the
same order as the flow time, ensuring that the flow will not only cool
by adiabatic expansion, but also radiatively, thus giving rise to the
fountain return flow, which is the part of the outflow that loses pressure
support from below and therefore cannot escape. Note that $r_c$ is the maximum
extension of the fountain flow in steady state. Thus, the lack of such an extended
$z$-grid inhibits the disk-halo-disk circulation of matter, which
otherwise would return gas to the disk sometime later, with noticeable
effects for the dynamical evolution there. The loss of matter may be compensated by some injection of mass by means of rather artificial {\bf boundary conditions} imposed into the upper and bottom faces of the grid.

The {\bf resolution} adopted in a simulation results from a compromise between
the available computing resources and the minimum scale required to handle
appropriately the physical processes involved in the system under study. For
instance in the ISM simulations high resolutions are required due to the
formation of small scale structures, resulting from instabilities in the flows, in particular from thermal instabilities and condensations as a result of radiative cooling. The amount of cooling may be substantially increased if the resolution is high enough to trace regions of high compression by
shocks, rather than smearing them out over larger cells, and thereby wiping out density peaks. Radiative cooling as a nonlinear process can become more
efficient, since high density regions contribute more to the energy
loss rate than low density regions can compensate by an accordingly
lower rate. Since cooling is most efficient for dense gas, the cool
phase is affected most. In addition, the spatial resolution of shear
layers and contact surfaces, gives rise to an increased level of
turbulence and a larger number of mixing layers.  The latter is most
important, because it allows for a faster mixing between parcels of
gas with different temperatures (conduction or diffusion processes
being of second order and hence inherently slow in nature). The small
scale mixing is promoted by numerical rather than molecular diffusion, and
therefore, the time scales for mixing in the different phases to occur is
somewhat smaller (because it happens on larger scales) than those predicted by molecular diffusion theory (e.g., Avillez \& Mac Low 2002). However, turbulent diffusion, as a consequence of the onset of turbulence due to shear flows, will be most efficient.
\begin{wrapfigure}[21]{r}[0pt]{7cm}
\centerline{\includegraphics[width=0.9\hsize,angle=0]
{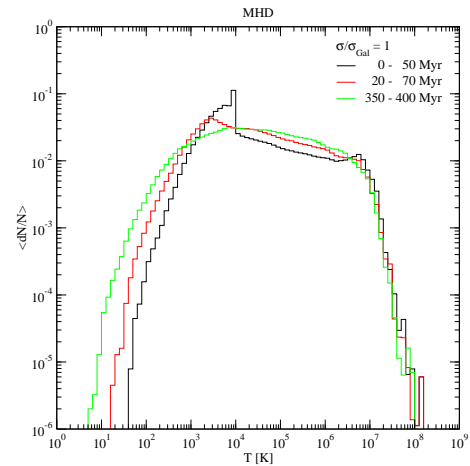}}
\caption{Averaged volume-weighted temperature histogram for a MHD global ISM simulation over the periods of 0-50 Myr (black), 20-70 Myr (red) and 350-400 Myr (green) calculated using 51 snapshots
taken at time intervals 1 Myr. The resolution of the finest AMR level is
1.25 pc.}
\label{resolution}
\end{wrapfigure}
\indent A necessary, but not sufficient, condition on {\bf convergence} of the simulations is that by increasing the resolution (normally by doubling it a few times) global properties, such as volume filling factors, history of minimum temperature and maximum density, mass distribution with time, etc, do not change significantly. When the numerical solutions, with the increase in resolution, have a small discrepancy of a few percent then one can safely conclude that the simulations converge for the  physical processes included. For further discussion see Avillez \& Breitschwerdt (2004). 

The initial evolution of a system imprints its signature in the averaged histograms of time evolved variables until the nonlinear processes developing during this time wipe out the signature of the initial conditions. In the case of global models as the ones discussed below, the initial evolution imprint is still seen after 70 Myr of evolution of a magnetized ISM (Fig.~\ref{resolution}, which shows averaged volume weighted histograms of the temperature over the periods of 0-50, 20-70 and 350-400 Myr for an MHD run with the Galactic SN rate and a finest AMR resolution of 1.25 pc). This imprint is also seen in corresponding histograms of HD simulations (Avillez \& Breitschwerdt 2004).

\section{Results from the Latest HD and MHD Simulations of a Disk-halo-disk Circulation System}

Taking into account the above considerations we carried out high resolution (using adaptive mesh refinement with finer resolutions ranging from 0.625 to 2.5 pc and coarse grid resolution of 10 pc) three-dimensional kpc-scale HD and MHD simulations of the ISM including the disk-halo-disk circulation with a grid centred on the solar circle and extending from $z=-10$ to 10 kpc with a square disk area of 1 kpc$^{2}$. The basic processes included are the gravitational field provided by the stars in the disk, radiative cooling assuming an optically thin gas in collisional ionization equilibrium, uniform heating due to starlight varying with $z$, supernovae (with a canonical explosion energy of $10^{51}$ erg) types Ia (with scale height, distribution and rate taken
\begin{wrapfigure}[48]{l}[0pt]{6cm}
\centering
\vspace*{11cm}
Jpeg image fig2.jpg
\vspace*{11cm}

\caption{Slice through the 3D data set showing
the vertical (perpendicular to the midplane) distribution of the
density at time 166 Myr.
\label{dh}
}
\end{wrapfigure}
from the literature) and Ib and II, whose formation,
and spatial location is calculated self-consistently by determining the number and masses and main sequence lifetimes of the OB stars formed in regions with $n\geq 10$ cm$^{-3}$, T$\leq 100$~K, respectively, and $\nabla\cdot \vec{v}<0$, where $\vec{v}$ is the gas velocity. Some $40-50\%$ of the stars (these are mainly the stars with masses $\leq 11\msolar$) explode in the field (the rest in associations) and have an average scale height of $\sim 90$ pc. The mean total rate per unit volume of occurrence of SNe in these simulations is $18~\mbox{kpc}^{-3}~ \mbox{Myr}^{-1}$, a value similar to the observed one. In the case of the MHD run the ISM is pervaded by a magnetic field with a total strength of $\simeq4.45~\mu$G resulting from unform and random components with mean values of 3.25 and $3.1~\mu$G, respectively.

\subsection{Global Evolution}

The simulations depart from an hydrostatic setup of the disk and halo gas
which does not hold for long as a result of the lack of equilibrium between
gravity and (thermal, kinetic and turbulent) pressure during the ``switch-on
phase'' of SN activity. As a consequence the gas in the upper and
lower parts of the grid collapses onto the midplane, leaving low
density material behind. However, in the MHD run it takes a longer
time for the gas to descend towards the disk and the complete collapse
into the midplane is prevented by the opposing magnetic pressure and
tension forces. As soon as enough SNe have gone off in the disk
building up the required pressure support, a thick frothy disk is
formed overlayed by a hot halo. The thick gas disk is feeded and supported by
the motions of the hot gas warmed up by randomly distributed SNe which rises
buoyantly and eventually breaks through into the halo. Transport into the halo
is not prevented, although the escape of the gas takes a few
tens of Myr to occur in the MHD run -- somewhat longer than in the
pure hydrodynamical case. The crucial point is that the huge thermal
overpressure due to combined SN explosions can sweep the magnetic field into
dense filaments and punch holes into the extended warm and ionized layers
allowing the setup of pressure release valves, after which there is no way from keeping the hot over-pressured plasma to follow the pressure and density gradient into the halo.

Fig.~\ref{dh} shows the density distribution in the
plane perpendicular to the Galactic midplane at time 166 Myr. Red/blue in the colour scale refers to lowest/highest density (or highest/lowest temperature). The $z-$scale above 0.5 and below -0.5 kpc is shrunk (in order to fit the paper size) and thus, the distribution of the labels is not uniform.  The image shows the presence of a wiggly thin disk of cold gas overlayed by a frothy thick disk (punctured by chimneys and crossed by hot buoyant ascending gas) composed of neutral (light blue), with a scale height of $\sim180$~pc, and ionized (greenish) gas with a scale height of
1~kpc. These distributions reproduce those described in Dickey \&
Lockman (1990) and Reynolds (1987), respectively. The upper parts of
the thick ionized disk form the disk-halo interface located around 2 kpc above and below the midplane, where a large
scale fountain is set up by hot ionized gas, injected there either from gas
streaming out of the thick disk or directly from superbubbles in the disk
underneath, escaping in a turbulent convective flow. The corresponding magnetic field maps, presented in Avillez \& Breitschwerdt (2005), show the presence of a thin magnetized disk overlayed by Parker-like loops (produced without cosmic rays), magnetic islands, and clouds wrapped in field lines moving downwards. There is also cold gas descending along the Parker loops. 
\begin{figure}[thbp]
\centering
\includegraphics[width=0.4075\hsize,angle=0]
{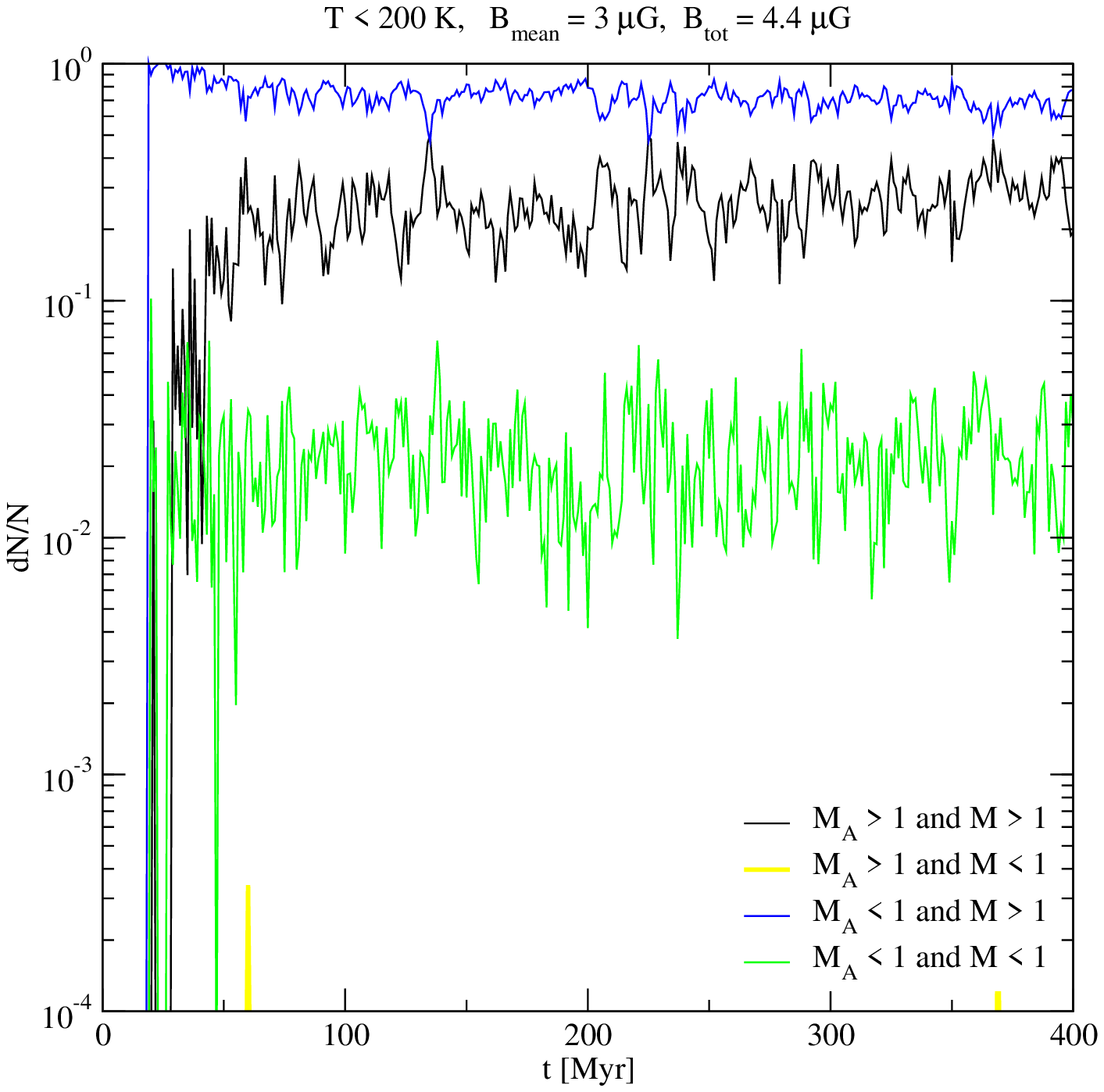}\hspace*{2cm}Jpeg image 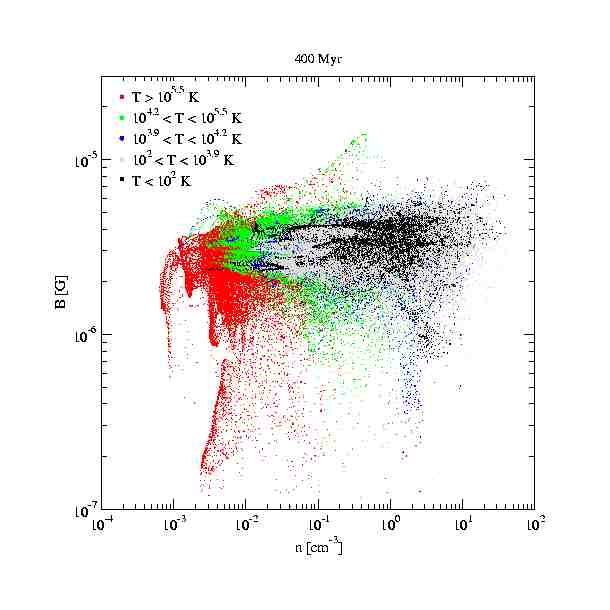
\caption{\emph{Left panel:} History of the cold ($\temp \leq 200$ K) gas fraction that is super/sub-alfv\'enic (denoted by $\mbox{M}_{A}> or < 1$) and
super/sub-sonic (denoted by $\mbox{M}> or < 1$). As it can be seen the
less than $10^{-2}$\% of cold gas has $\mbox{M}_{A}>1$ and $\mbox{M}<1$. \emph{Right panel:} Scatter plot of B versus $n$ for $\temp\leq 10^{2}$
  (black), $10^{2}<\temp\leq 10^{3.9}$ (grey), $10^{3.9}<\temp\leq
  10^{4.2}$ (blue), $10^{4.2}<\temp\leq10^{5.5}$ K (green), and
  $\temp>10^{5.5}$ K (red) regimes at 400 Myr of disk evolution. The
  points in the plot are sampled at intervals of four points in each
  direction. 
\label{alfvenic}
\label{scatterDB}
}

\end{figure}
\subsection{Shock Compressed Layers}

The highest density gas with $\temp \leq 10^{2}$ K is confined
to shocked compressed layers that form in regions where
several large scale streams of convergent flow (driven by SNe)
occur. The compressed regions, which have on average lifetimes of a
few free-fall times, are filamentary in structure, tend to be aligned
with the local field and are associated with the highest field
strengths (in the MHD run), while in the HD runs there is no
preferable orientation of the filaments. The formation time of these
structures depends on how much mass is carried by the
convergent flows, how strong the compression and what the rate of
cooling of the regions under pressure are. During the dynamical
equilibrium evolution on average 70\% of the cold ($\temp \leq$
200 K) gas is subalfv\'enic and supersonic, while only 1-5\% is
subalfv\'enic and subsonic, the remaining fraction of the gas has
$\mbox{M}_{A}>1$ and $\mbox{M}>1$ (Fig.~\ref{alfvenic}). This means 
that although in the ISM the majority of the cold gas is
supersonic and subalfvenic there is a considerable fraction ($\langle dN/N\rangle \simeq 30\%$) of the gas where the mean magnetic pressure is dynamically low, i.e., $\mbox{M}_{A}>1$.

\subsection{Field Dependence with Density}

After the global dynamical equilibrium has been set up the magnetic
field shows a high variability (which decreases towards
higher gas densities spanning two orders of magnitude from $0.1$ to $15$
$\mu$G; see Fig.~\ref{scatterDB}) and it is \emph{largely uncorrelated
with the density}. The spreading in the field strength
increases with temperature, being largest for the hot ($\temp >
10^{5.5}$ K) and smallest (with almost an order of magnitude
variation from 0.8 to 6 $\mu$G) for the cold ($\temp \leq 200$ K)
gas. The large scatter in the field strength for the \textit{same} density,
seen in Fig.~\ref{scatterDB}, suggests that the field is being
driven by the inertial motions, rather than it being the agent
determining the motions. In the latter case the field would not be
strongly distorted, and it would direct the motions predominantly
along the field lines. In ideal MHD field diffusion is negligible, and
the coupling between matter and field should be perfect (we are
of course aware of numerical diffusion which weakens the argument at
sufficiently small scales). Therefore gas compression is correlated with
field compression, except for strictly parallel flow. 

\subsection{Driving Forces in the ISM}
\begin{wrapfigure}[19]{r}[0pt]{7cm}
\centerline{\includegraphics[width=0.9\hsize,angle=0]{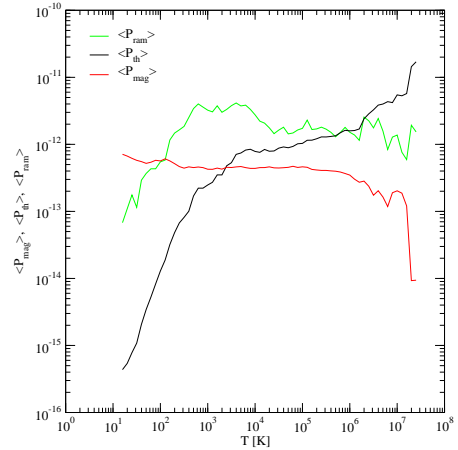}
}
\caption{Average $\langle \pram \rangle$ (green),
$\langle \pth \rangle$ (black), and $\langle \pmag \rangle$ (red)
as function of the temperature (in the simulated disk
$\left|z\right|\leq 250$ pc) averaged over temperature bins of $\Delta
\log \temp=0.1$ K  between 350 and 400 Myr.
\label{rampressure2}
}
\end{wrapfigure}
The relative importance of the driving pressures, i.e., thermal ($\pth$), ram
($\pram$) and magnetic ($\pmag$), in the ISM varies with increasing temperature
(Fig.~\ref{rampressure2}).  For $\temp \leq 150$ K $\pth<<\pram<\pmag$
indicating that the gas is dominated by the Lorentz $\vec{j} \times \vec{{\rm
B}}$ force, and the magnetic field determines the motion of the fluid, while for
$\temp \geq 10^{5.5}$ K thermal pressure dominates. At the
intermediate temperatures ram pressure determines the dynamics of the
flow, and therefore, the magnetic pressure does not act as a
significant restoring force (see Passot \& V\'azquez-Semadeni 2003) as
it was already suggested by the lack of correlation between the field
strength and the density. It is also noteworthy that in this
temperature range ($150 - 10^{5.5}$ K) the weighted magnetic pressure
is roughly constant (suggesting that the magnetic and thermal
pressures are largely independent, whereas both thermal and ram
pressures undergo large variations in this temperature interval.
\subsection{The Chandrasekhar-Fermi Law}
As ram pressure fluctuations in the ISM dominate over the other
pressures a perfect correlation between the field and density following the classical scaling law $B\sim\rho^\alpha$ is not expected, with $\alpha=1/2$,
according to the Chandrasekhar-Fermi (CF) model (1953), but a broad distribution
of $B$ versus $\rho$. Although in general $0\leq \alpha \leq 1$ would be
expected, it should be noted that in reality heating and cooling
processes, and even magnetic reconnection could induce further
changes, making the correlation rather complex. It should
be kept in mind that in CF it was assumed that the field is distorted
by turbulent motions that were subalfv\'enic, whereas in our
simulations in addition both supersonic and superalfv\'enic motions
can occur, leading to strong MHD shocks. In other words, according to
the CF model, the turbulent velocity is directly proportional to the
Alfv\'en speed, which in a SN driven ISM need not be the case.

\subsection{Pressure Distributions and Fluctuations}

\begin{wrapfigure}[21]{r}[0pt]{7cm}
\centerline{\includegraphics[width=0.9\hsize,angle=0]
{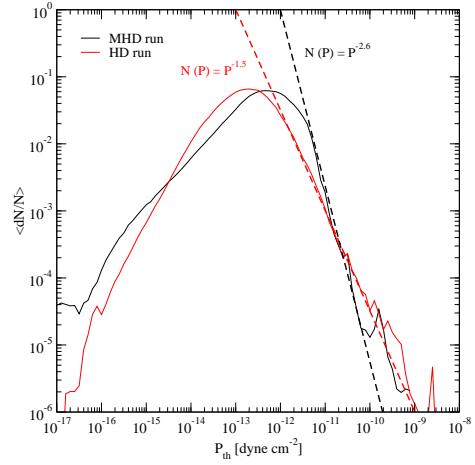}}
\caption{Averaged thermal pressure distribution in the simulated disk
($\left|z\right|\leq 250$ pc) for the MHD (black) and HD (red) runs. These pdfs are calculated by using 51 snapshots taken between 350 and 400 Myr with a time interval of 1 Myr. The right side of the histograms is overlayed with a straight dashed line corresponding to power laws with negative slopes 1.5 and 2.6.
\label{mhdpdfs}
}
\end{wrapfigure}

The pressure coverage of three orders of magnitude, for volume fractions of dN/N$\geq 10^{-2}$, seen in the averaged (over the period 350-400 Myr) volume weighted histograms of $\pth$ (Fig.~\ref{mhdpdfs}) is similar for the magnetized and unmagnetized ISM runs, although the power law fits to their profiles have different negative slopes: 2.6 and 1.5, respectively. These results are indicative of the large fluctuations in thermal pressure between
different temperature regimes, suggesting that there are no real "phases", i.e., co-existing thermodynamic regimes with different density and temperature but in pressure equilibrium (see also Avillez \& Breitschwerdt 2004, 2005; Mac Low et al. 2005).
\subsection{Volume Weighted Histograms}
A comparison between the averaged (over the time 350 through 400
Myr) volume weighted histograms of the density and temperature of
the magnetized and unmagnetized disk gas shows differences that include the (i) decrease/increase by an almost order of magnitude in the histograms
density/temperature coverage, (ii) change in the relative weight of the dominant temperature regimes (in the density histograms) and consequently
(iii) changes in the pronounced bimodality of the total density and
temperature histograms. This latter effect is most noticeable in the
temperature PDFs (see Fig.~\ref{resolution}). In effect, while the
temperature PDF of the HD run has a bimodal structure (as it has two
peaks: one at 2000 K and another around $10^{6}$ K), in the MHD run
the decrease/increase in importance of the $10^{3.9} <\temp
\leq 10^{4.2}$ K/$10^{4.2} <\temp \leq 10^{5.5}$ K leads to the
reduction/increase of the occupation fraction of these regimes and
therefore to a change of the histogram structure appearing it to be
unimodal. This variation of the intermediate region appears to be an
effect of the presence of the magnetic field, with the smoothing
effect being less pronounced for lower field strengths in the disk.

\subsection{Volume Filling Factors}
During most of the history ($t> 100$ Myr) of ISM
evolution the occupation ($\mbox{f}_{\mbox{v}}$) fractions of the
different thermal regimes have an almost
constant distribution, varying around their mean
values (cf.~Table 1; see also Avillez 2000; Avillez \& Breitschwerdt 2004,
2005). The thermally stable regimes with $\temp\leq
200$ K and $10^{3.9}<\temp \leq 10^{4.2}$ K have similar occupation
fractions of $\sim 5\%$ and $\sim 10\%$, respectively, in both runs,
while the hot gas has an increase from $\sim 17\%$ in the HD run to
$\sim 20\%$ in the MHD case. By far the disk volume is occupied by
gas in the thermally unstable regimes at $200<\temp \leq 10^{3.9}$
and $10^{4.2}<\temp \leq 10^{5.5}$ K with similar occupation
fractions $\sim 30\%$ in the MHD run, while in the HD run these
regimes occupy $46\%$ and $22\%$, respectively, of the disk volume. 

These results indicate that the presence of the magnetic field,
which may inhibit the break-out of an individual remnant, but
certainly not the high-pressure flow resulting from supernova explosions in
concert within an OB association, only leads to a slight increase in the
occupation fraction of the hot gas in the disk. The reason is that the volume filling factor of the hot gas  increases slightly is because magnetic
tension forces help to confine bubbles and Lorentz forces obstruct mixing
with cooler gas. It is plausible to assume that, similarly to what seen in
HD simulations with different star forming rates, there is a correlation between
the filling factor of the hot gas and the rate at which SN occur, since higher
rates will produce more hot plasma which, as we have shown here, is not
magnetically controlled.

\begin{table}[thbp]
\centering \caption{Summary of the average values of volume filling
factors, mass fractions and root mean square velocities of the disk
gas at the different thermal regimes for the HD and MHD
runs (from Avillez \& Breitschwerdt 2005).\label{table1}
}
\begin{tabular}{c|cc|cc|cc}
\hline
\hline
T  & \multicolumn{2}{c|}{$\langle \mbox{f}_{\mbox{v}}\rangle^{a}$ [\%]}&
\multicolumn{2}{c|}{$\langle \mbox{f}_{\mbox{M}}\rangle^{b}$ [\%]} &
\multicolumn{2}{c}{$\langle \mbox{v}_{\mbox{rms}}\rangle^{c}$}\\
\cline{2-7}
[K] & HD & MHD & HD & MHD & HD & MHD\\
\hline
$<200$ K              & ~5 & ~6 &  44.2 & 39.9 & ~7 & 10 \\
$200-10^{3.9}$        & 46 & 29 &  49.0 & 43.7 & 15 & 15 \\
$10^{3.9}-10^{4.2}$   & 10 & 11 &  ~4.4 & ~8.5 & 25 & 21 \\
$10^{4.2}-10^{5.5}$   & 22 & 33 &  ~2.0 & ~7.4 & 39 & 28 \\
$>10^{5.5}$           & 17 & 21 &  ~0.3 & ~0.5 & 70 & 55 \\
\hline
\multicolumn{7}{l}{$^a$ volume filling factor.}\\
\multicolumn{7}{l}{$^b$ Mass fraction.}\\
\multicolumn{7}{l}{$^c$ Root mean square velocity in units of km s$^{-1}$.}
\end{tabular}
\end{table}

\subsection{ISM Mass Fractions and Warm Neutral Medium Mass}
\begin{wrapfigure}[19]{l}[0pt]{7cm}
\centerline{\includegraphics[width=0.9\hsize,angle=0]
{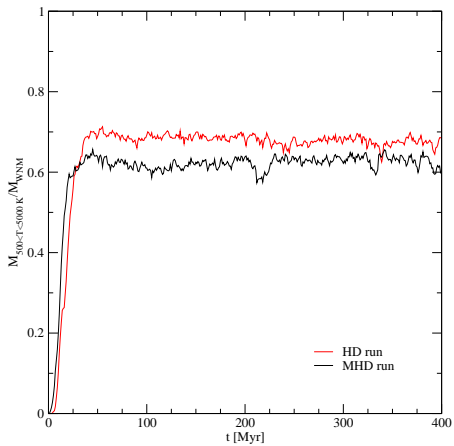} }
\caption{History of the fraction of mass of the WNM gas having $500
<\temp \leq 5000$ K in the disk for the HD (red) and MHD (black) runs.
\label{wnmmass}
}
\end{wrapfigure}
Most of the disk mass is found in the $\temp\leq 10^{3.9}$ K gas,
with the cold ($\temp \leq 200$ K) and thermally unstable gases
($200<\temp\leq 10^{3.9}$ K) harbouring on average 80 and $\sim 90\%$ of
the disk mass in the MHD and HD runs (Table~\ref{table1}), with the
cold regime enclosing $\sim 40\%$ of the disk mass. The remaining
ISM mass is distributed between the other temperature regimes with the
$10^{3.9} <\temp\leq 10^{4.2}$ K and $10^{4.2} <\temp\leq 10^{5.5}$ K
regimes enclosing a total of only $\sim 7\%$ and $\sim 16\%$ of mass in the
HD and MHD runs, respectively, and the hot gas enclosing $<1\%$ of the
disk mass in both runs. In both runs, 60-70\% of the warm neutral mass
($500<\temp \leq 8000$ K) is contained in the $500\leq\temp\leq 5000$ K
temperature range (bottom panel of Fig.~\ref{wnmmass}). 

This latter result is strongly supported by interferometric (Kalberla et al. 1985) and optical/UV absorption-line measurements (Spitzer \& Fitzpatrick 1995, Fitzpatrick \& Spitzer 1997), which indicate that a large fraction ($\sim 63\%$) of the warm neutral medium (WNM) is in the unstable range
$500<\temp<5000$ K. Moreover 21~cm line observations (Heiles 2001;
Heiles \& Troland 2003) provide a lower limit of $48\%$ for the WNM
gas in this unstable regime. Direct numerical simulations of the
nonlinear development of the thermal instability under ISM
conditions with radiative cooling and background heating discussed
in Gazol et al. (2001) and Kritsuk \& Norman (2002) show that about
60\% of the system mass is in the thermally unstable regime.
However, it is unclear from their simulations what the time
evolution of this mass fraction (which is shown in the figure for $t> 100$ Myr) is and what explicitly the origin of
the unstable gas is. These authors suggest that ensuing
turbulence is capable of replenishing gas in the thermally unstable
regime by constantly stirring up the ISM.  We have carried out
detailed numerical studies of the stability of the ISM gas phases
(Avillez \& Breitschwerdt 2005a), and verified the hypothesis
that SN driven turbulence is capable of replenishing fast cooling
gas in classically unstable regimes. In effect, turbulence as a diffusion process can prevent thermal runaway on small scales. That is, turbulence has a stabilizing effect thereby inhibiting local condensation modes.The turbulent viscosity $\nu_{\rm
turb} \sim Re \, \nu_{\rm mol}$ can be orders of magnitude above the
molecular viscosity, with $Re$ being the Reynolds number of the
flow.  What happens physically then, is that with increasing eddy
wavenumber $k=2\pi/\lambda$, the eddy crossing time $\tau_{\rm eddy}
\sim \lambda/\Delta u$ (with $\Delta u$ being the turbulent velocity
fluctuation amplitude) becomes shorter than the cooling time
$\tau_{\rm cool} \sim 3 k_B T/(n \Lambda(T)) $, where $\Lambda(T)$ is
the interstellar cooling function. Although not strictly applicable
here, it is instructive to see that in case of incompressible
turbulence following a Kolmogoroff scaling law, where the energy
dissipation rate is given by $\varepsilon \sim
\rho {\Delta u}^3/\lambda$, we obtain a lower cut-off in wavelength,
where thermal instability becomes inhibited, if
\begin{eqnarray}
  \lambda  &<&  \left(\frac{3 k_B \bar
m}{\Lambda_0} \right)^{3/2} \, \varepsilon^{1/2} \, \frac{T^{3/4}}{\rho^2}\\
  &\approx & 1.42 10^{19} \, {\rm cm} \nonumber \,,
  \label{lambdacut}
\end{eqnarray}
taking $\varepsilon \sim 10^{-26}\, {\rm erg} \, {\rm cm}^{-3}\,
{\rm s}^{-1}$ for SN energy injection; a simple cooling law for the
warm neutral medium of $\Lambda(T) = \Lambda_0 T^{1/2}$ has been
adopted, with $\Lambda_0 \approx 1.9 \times 10^{-27}\, {\rm erg} \,
{\rm cm}^3 \, {\rm s}^{-1} \, {\rm K}^{-1/2}$ (taken from the
cooling curve of Dalgarno \& McCray 1972) for a WNM of a density of
$n = 0.3 \, {\rm cm}^{-3}$, a temperature of $T=1000$ K, and a low
degree of ionization $x \approx 0.01$. Therefore rough numerical
estimates are typically of the order of parsecs, consistent with our
numerical resolution.  In fact, the critical wavelength $\lambda$
varies with temperature, degree of ionization and hence cooling; for
the WNM we find quite a large range of values from $10^{17} -
10^{20}$ cm, according to Eq.~(1). 

\subsection{Turbulent Velocities}

The root mean square velocity, $\vrms$, which is a measure of the
disordered motion of the gas, increases with temperature
(see Table~\ref{table1}) in the MHD and HD runs. The average rms velocity
($\langle\vrms\rangle$) in the last 100 Myr of evolution has large fluctuations in the different thermal regimes in the HD run, which are reduced due to the presence of the magnetic field. These velocities agree remarkably well with the observed rms velocities discussed by Kulkarni \& Fich (1985). The near constancy of the rms velocity with time indicates the presence of a dynamical equilibrium, with random motions, i.e.\ thermal and turbulent pressures adding to the total pressures, provided that the energy
injection rate remains constant on a global scale.

\section{Testing the Model}

Although the above mentioned results are in good agreement with present day observations of the ISM and other numerical experiments, though these have more a local character, one can trace the quality of the present model by applying it to the dynamics and evolution of the Local Bubble, powered by the explosions of 19 SNe in the last 14 Myr (Bergh\"ofer \& Breitschwerdt 2002; Fuchs et al. 2005), and trace its O{\rm VI} content that has been observed using Copernicus and FUSE. While standard Local Bubble (LB) models fail to reproduce the observed low O{\rm VI} absorption column density (Shelton \& Cox 1994; for a recent discussion see Breitschwerdt \& Cox 2004), the present model when applied to the study of the dynamics and evolution of the Local and Loop I bubbles predicts column densities $<1.7 \times 10^{13} \, {\rm cm}^{-2}$ towards Loop I with a mean value of $\leq 8.5\times 10^{12}$ cm$^{-2}$ (for details see Breitschwerdt et al. these proceedings) in agreement with the mean column density of $7\times 10^{12}$ cm$^{-2}$ inferred from analysis of \textsc{FUSE} absorption line data in the Local ISM (Oegerle et al. 2004).

\section{Final Remarks}

The present simulations still neglect an important component of the
ISM, i.e., high energy particles, which are known to be in rough
energy equipartition \textit{locally} with the magnetic field, the
thermal and the turbulent gas in the ISM. The presence of CRs and
magnetic fields in galactic halos is well known and documented by
many observations of synchrotron radiation generated by the electron
component. The fraction of cosmic rays that dominates their total energy is of Galactic origin and can be generated in SN remnants via the
diffusive shock acceleration mechanism to energies up to $10^{15}$
eV (for original papers see Krymski et al. 1977, Axford et al. 1977,
Bell 1978, Blandford \& Ostriker 1978, for a review Drury 1983, for
more recent calculations see Berezhko 1996). The propagation of these particles generates MHD waves due to the streaming instability (e.g.\ Kulsrud \& Pearce
1969) and thereby enhances the turbulence in the ISM. In addition,
self-excited MHD waves will lead to a dynamical coupling between the
cosmic rays and the outflowing fountain gas, which will enable part
of it to leave the galaxy as a galactic wind (Breitschwerdt et al.
1991, 1993, Dorfi \& Breitschwerdt 2005). Furthermore, as the cosmic rays act as a weightless fluid, not subject to radiative cooling, they can bulge out magnetic field lines through buoyancy forces. Such an inflation of the field will inevitably lead to a Parker type instability, and once it becomes nonlinear, it will break up the field into a substantial component parallel to the flow (Kamaya et al.\ 1996), thus facilitating gas outflow into the halo. We are currently performing ISM simulations including the CR component.

In the dynamical picture of the ISM emerging from our simulations,
thermal pressure gradients dominate mainly in the neighbourhood of
SNe. These events drive motions whose ram pressures overwhelm the
mean thermal pressure (away from the energy sources) and the
magnetic pressure by a large factor. The magnetic field is
dynamically important at low temperatures, apart from also weakening
gas compression in MHD shocks and thereby lowering the energy
dissipation rate. The thermal pressure of the freshly shock heated
gas exceeds the magnetic pressure by usually more than an order of
magnitude and the B-field can therefore not prevent the flow from
rising perpendicular to the galactic plane. Thus hot gas is fed into
the galactic fountain at almost a similar rate than without field
(Avillez \& Breitschwerdt 2004, 2005). 

However, the circulation of gas between the disk and halo is a dynamic process, which involves a flow time scale, that can be much shorter than any of the microphysical time scales due to ionization and recombination. The gas escaping into the halo has an initial temperature well in excess of $10^{6}$ K, where the assumption of collisional ionization equilibrium (CIE) is approximately valid. This means that the rate of ionizations per unit volume through collisions between ions and electrons equals the rate of recombinations per unit volume. As the hot plasma expands away from the disk it will cool adiabatically thereby reducing its temperature and density. It has been shown (Breitschwerdt \& Schmutzler 1999) that recombination of highly ionised species lags behind and occurs mainly at considerable heights from the disk. In the case of the X-ray halo of NGC 3029, spectral fitting has shown that an outflow model based on non-equilibrium ionization (NEI) effects gives an excellent agreement with observations (see Breitschwerdt 2003).

\section*{Acknowledgments}
MA would like to thank the SOC and Thierry Montmerle and Almas Chalabaev for the invitation and financial help to attend this excellent conference.

\end{document}